\documentclass[twocolumn,showpacs,preprintnumbers,amsmath,amssymb]{revtex4}
\usepackage{graphics}
\usepackage{epsfig}
\usepackage{bm}

\begin{document}
\title{Lattice dynamics in magnetic superelastic Ni-Mn-In alloys. Neutron
scattering and ultrasonic experiments}

\author{Xavier Moya\cite{Address}, David Gonz\'{a}lez-Alonso, Llu\'{i}s
Ma\~nosa}
\email{lluis@ecm.ub.es}
\author{and Antoni Planes}

\affiliation{Departament d'Estructura i Constituents de la
Mat\`eria, Facultat de F\'isica, Universitat de Barcelona,
Diagonal 647, E-08028 Barcelona, Catalonia, Spain}

\author{V. O. Garlea, T. A. Lograsso, D. L. Schlagel and J. L. Zarestky}
\affiliation{Ames Laboratory, Iowa State University, Ames, Iowa 50011}

\author{Seda Aksoy and Mehmet Acet}
\affiliation{Fachbereich Physik, Experimentalphysik,
Universit\"{a}t Duisburg-Essen, D-47048 Duisburg, Germany}
\date{\today}

\begin{abstract}

Neutron scattering and ultrasonic methods have been used to study
the lattice dynamics of two single crystals of Ni-Mn-In Heusler
alloys close to Ni$_{50}$Mn$_{34}$In$_{16}$ magnetic superelastic
composition. The paper reports the experimental determination of
the low-lying phonon dispersion curves and the elastic constants
for this alloy system. We found that the frequencies of the
TA$_{2}$ branch are relatively low and it exhibits a small dip
anomaly at a wave number $\xi_{0} \approx 1/3$, which softens with
decreasing temperature. Associated with the softening of this
phonon, we also observed the softening of the shear elastic
constant $C'=(C_{11}-C_{12})/2$. Both temperature softenings are
typical for bcc based solids which undergo martensitic
transformations and reflect the dynamical instability of the cubic
lattice against shearing of $\{110\}$ planes along $\langle
1\bar{1}0 \rangle$ directions. Additionally, we measured low-lying
phonon dispersion branches and elastic constants in applied
magnetic fields aimed to characterize the magnetoelastic coupling.

\end{abstract}


\maketitle

\section{Introduction}

The search of magnetic shape memory alloys with more favourable mechanical
properties than Ni$_{2}$MnGa has prompted in recent years the study of
new Ni-Mn based Heusler alloys, extending this alloy family to other elements
of groups IIIA-VA \cite{Sutou2004,Krenke2005a,Krenke2006,Krenke2005b}.
Among them, Ni-Mn-In system has drawn much attention due to the large shift of
the martensitic transition temperatures by an applied magnetic field
observed in Ni$_{50}$Mn$_{34}$In$_{16}$ ($\sim 10$ K/T), as a consequence of a
strong spin-lattice coupling at a microscopic level
\cite{Krenke2007}. These large shifts allow for the application of moderate
magnetic fields to induce the structural transition and lead to many
functional properties such as the magnetic superelasticity \cite{Krenke2007},
large inverse magnetocaloric effects
\cite{Han2006,Moya2007,Planes2009} and magnetoresistance \cite{Sharma2006}.

From a fundamental point of view, many of the Ni-Mn based alloys
exhibit singular lattice-dynamical behavior associated with the
martensitic transition from a high temperature cubic phase to a
lower symmetry martensitic phase. Specifically, in the case of
Ni-Mn-Ga and Ni-Mn-Al alloys, it has been experimentally shown
that the transverse TA$_{2}$ branch shows a dip at a particular
wave number (anomalous phonon). The energy of such anomalous
phonon softens with decreasing the temperature
\cite{Zheludev95,Zheludev96,Stuhr97,Manosa01,Moya2006a,Mehaddene08}.
The temperature dependence of the energy of the anomalous phonon
parallels that of the elastic constant $C'=(C_{11}-C_{12})/2$
which also softens with decreasing temperature
\cite{Worgull96,Manosa97,Stipcich04,Moya2006b}. Softening observed
both with neutron scattering and ultrasonic methods are typical
for bcc based solids which undergo martensitic transformations and
reflect the dynamical instability of the cubic lattice against
shearing of $\{ 110\}$ planes along the $\langle 1\bar{1}0
\rangle$ directions \cite{Planes2001}. In addition, significant magnetoelastic
coupling exists in these systems as evidenced by the enhancement
of the anomalous phonon softening when the sample orders
ferromagnetically \cite{Stuhr97,Manosa01}, and by the change in
the elastic constants when a magnetic field is applied
\cite{GonzalezComas99,Moya2006b}.

The study of the lattice dynamics of a broader class of Ni-Mn
based compounds is important for a deeper understanding of the
physical mechanisms behind the multifunctional properties of
martensitic Heusler alloys. Therefore, much effort has been
devoted in recent years to extend the study of the lattice
dynamics to alloys other than Ni-Mn-Ga and Ni-Mn-Al by means of
first principles calculations \cite{Zayak05,Entel2008}. In
particular, Entel \textit{et al.} have shown that the application
of a magnetic field leads to the gradual vanishing of the phonon
instability in the low-lying TA$_{2}$ branch for stoichiometric
Ni$_{2}$MnIn, thus stabilizing the high temperature cubic phase
\cite{Entel2008}. The aim of the present work is to extend the
experimental study of the lattice dynamics of this alloy-family to
the Ni-Mn-In system, both in the short-wavelength and the
long-wavelength limit by measuring the phonon dispersion branches
and the elastic constants. Due to the strong interplay between
magnetic and structural degrees of freedom typically exhibited by
these compounds, we carried out both kinds of experiments in
applied magnetic field.

The paper is organized as follows: In Section
\ref{sec:ExperimentalDetails} we describe the details of single
crystal preparation and experimental techniques. Section \ref{sec:Results} is
devoted to the experimental
results and is split into three sections, which describe the calorimetric
(\ref{subsec:Calorimetry}), phonon dispersion
(\ref{subsec:PhononDispersion}) and elastic constants
(\ref{subsec:ElasticConstants}) results. Finally, in Section
\ref{sec:Summary} we summarize and conclude.

\section{Experimental Details}
\label{sec:ExperimentalDetails}

\subsection{Sample preparation}

The single crystals were synthesized at the Materials Preparation
Center, Ames Laboratory, USDOE \cite{AmesLab}. Two single crystals
with similar compositions were prepared, one for neutron
scattering experiments and another one for ultrasonic experiments.
Appropriate quantities of high purity nickel, manganese and indium
were used to make these alloys. The electrolytic manganese was
pre-arc melted to outgas it before alloying it with the other
metals.  The metals were arc melted several times under an argon
atmosphere, flipping the buttons each time.  The buttons were then
remelted and the alloy drop cast into a copper chill cast mold to
ensure compositional homogeneity throughout the ingots.  The
crystals were grown in a resistance furnace from the as-cast
ingots in an alumina Bridgman style crucible. The ingots were
heated under a pressure of $5.0 \times 10^{-6}$ torr up to $1000$
$^{\circ}$C and $900$ $^{\circ}$C, for the neutron scattering and
ultrasonic method samples, respectively, to degas the crucible and
charge. The chamber was then backfilled to a pressure of $2.8
\times 10^{3}$ kPa with high purity argon.  This
over-pressurization was done near melting to diminish gas pockets
from being trapped in the cone region of the crystal and also to
minimize the amount of manganese evaporation from the melt during
crystal growth.  The ingots were further heated to $1250$
$^{\circ}$C and $1200$ $^{\circ}$C, for the neutron scattering and
ultrasonic measurements samples, respectively, and held at this
temperature for 1 hour to allow thorough mixing before withdrawing
the samples from the heat zone at a rate of 5 mm/hr. The as-grown
ingots were heat treated at $900$ $^{\circ}$C for 1 week and
cooled at a rate of $10$ $^{\circ}$C/min. The average composition
of the alloys were determined by energy dispersive X-ray analysis
(EDX) to be Ni$_{49.3}$Mn$_{34.2}$In$_{16.5}$ and
Ni$_{48.8}$Mn$_{34}$In$_{17.2}$ (within $\pm 0.5$ at. \%) for the
samples used in neutron scattering and ultrasonic experiments,
respectively.

\subsection{Calorimetric measurements}

Structural and magnetic transitions of the samples were characterized by means
of differential scanning calorimetry (DSC)
measurements. Calorimetric measurements were carried out in the
temperature range $150$ K $\leq T \leq 375$ K in a DSC (TA
Instruments MDSC 2920) at cooling and heating rates of 5 K/min. We
have also used a second high sensitivity calorimeter for
measurements in the temperature range $100$ K $\leq T \leq 350$ K.
In this case, typical cooling and heating rates were 0.5 K/min.

\subsection{Neutron scattering}

The crystals used in the experiments were cut from a large boule to a size
allowing
mounting in the various sample environment systems.  A side
benefit of this was a larger surface-to-volume ratio, important
because of the large absorption cross section of In for thermal
neutrons (194 bn). The samples were still sizeable ($\sim 4$ and
$\sim 6$ cm$^{3}$) although not perfect single crystals. Rocking
curves of the $(220)$ Bragg reflections used for sample
orientation showed, in one case, a secondary peak, which was very
close to the primary peak ($< 1^{\circ}$ away) and, in the other
case, a secondary peak with much lower intensity than the primary
peak ($\sim 20$ \%). Peak-widths of the $(220)$ rocking curves
were from $0.5^{\circ}$ to $1^{\circ}$. The secondary peaks
warranted caution in interpretation of the data but were not
prohibitive in performing the experiments.

The inelastic neutron scattering experiments were performed using
the HB1A triple-axis spectrometer at the High Flux Isotope Reactor
(HFIR) of the Oak Ridge National Laboratory (ORNL). The
monochromator and analyzer used the $(002)$ reflection of
pyrolitic graphite (PG). Highly oriented PG filters (HOPG) were
used to minimize higher-order contaminations of the beam. The HB1A
spectrometer operates at a fixed incident energy of 14.6 meV
requiring most scans to be performed with neutron-energy gain.
Nominal collimations of 48'-48'-40'-68' (or 136') were used and
all scans were performed in the constant-Q mode while counting
against neutron monitor counts.

The temperature dependent dispersion curve measurements were made
in a closed-cycle helium refrigerator (CCHR) with a high
temperature interface unit enabling the sample to be warmed to
temperatures above room temperature. Temperature control was
better than $\pm 1$ K and accuracy, due to temperature sensor
location, $\sim \pm 5$ K.  The magnetic field measurements were
made in a conventional superconducting cryomagnet with temperature
control again to within $\pm 1$ K.  The sample location on the
HB1A triple-axis spectrometer which is near the massive and
magnetic monochromator drum of the HB1 triple-axis spectrometer,
requires limiting applied magnetic fields to $\sim 4$ T.

\begin{figure}
\includegraphics[width=0.9\linewidth,clip=]{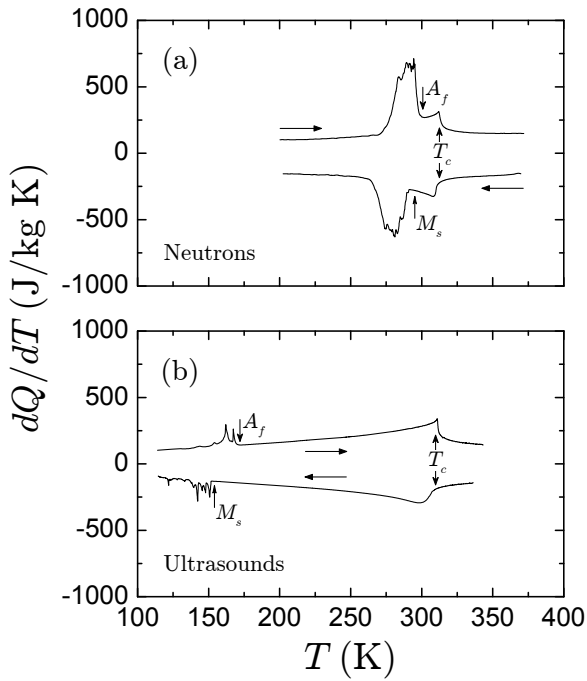}
\caption{Calorimetric curves for the samples used in (a) neutron
scattering (Ni$_{49.3}$Mn$_{34.2}$In$_{16.5}$) and (b) ultrasonic
measurements (Ni$_{48.8}$Mn$_{34}$In$_{17.2}$). Vertical arrows
show the position of martensitic and Curie transition
temperatures. Heating and cooling runs are shown with horizontal
arrows.} \label{fig1}
\end{figure}

\subsection{Ultrasonic methods}

The single crystal was oriented in the austenite phase by Laue back reflection,
and a parallelepiped
specimen with dimensions $5.0 \times 5.0 \times 3.7$ mm$^{3}$ with
faces parallel to the $(1\bar{1}0)$, $(110)$, and $(001)$ planes,
respectively, was spark cut from the oriented boule. The faces
were ground parallel and flat using standard metallographic
techniques.

The velocity of ultrasonic waves was determined by the pulse-echo
technique. X-cut and Y-cut transducers with resonant frequencies
of 10 MHz were used to generate and detect the ultrasonic waves.
The transducers were acoustically coupled to the surface of the
sample by means of Dow Corning Resin 276-V9 in the temperature
range 200--320 K and by Crystalbond509 (Aremco Products, Inc.) in
the temperature range 310--360 K. For high-temperature
measurements (up to 360 K), the sample was placed into a copper
sample holder which was heated by means of a heating plate. For
low-temperature measurements (down to 200 K), the sample and sample
holder were introduced into a dewar glass containing liquid
nitrogen. In both cases, the temperature was measured by a
Ni-Cr/Ni-Al thermocouple attached to the sample.

The measurements of the magnetic field dependence of the elastic
constants were carried out in a purpose-built device that allows
both isofield and isothermal measurements of the ultrasonic
velocities in applied fields up to 1.3 T (details are given in
reference \cite{Moya2006b}). The measurements were carried out
at constant temperature and the magnetic field was applied
perpendicular to the propagation direction of the ultrasonic
waves.

\section{Experimental results and discussion}
\label{sec:Results}

\subsection{\label{subsec:Calorimetry}Calorimetry}

\begin{figure}
\includegraphics[width=1.0\linewidth,clip=]{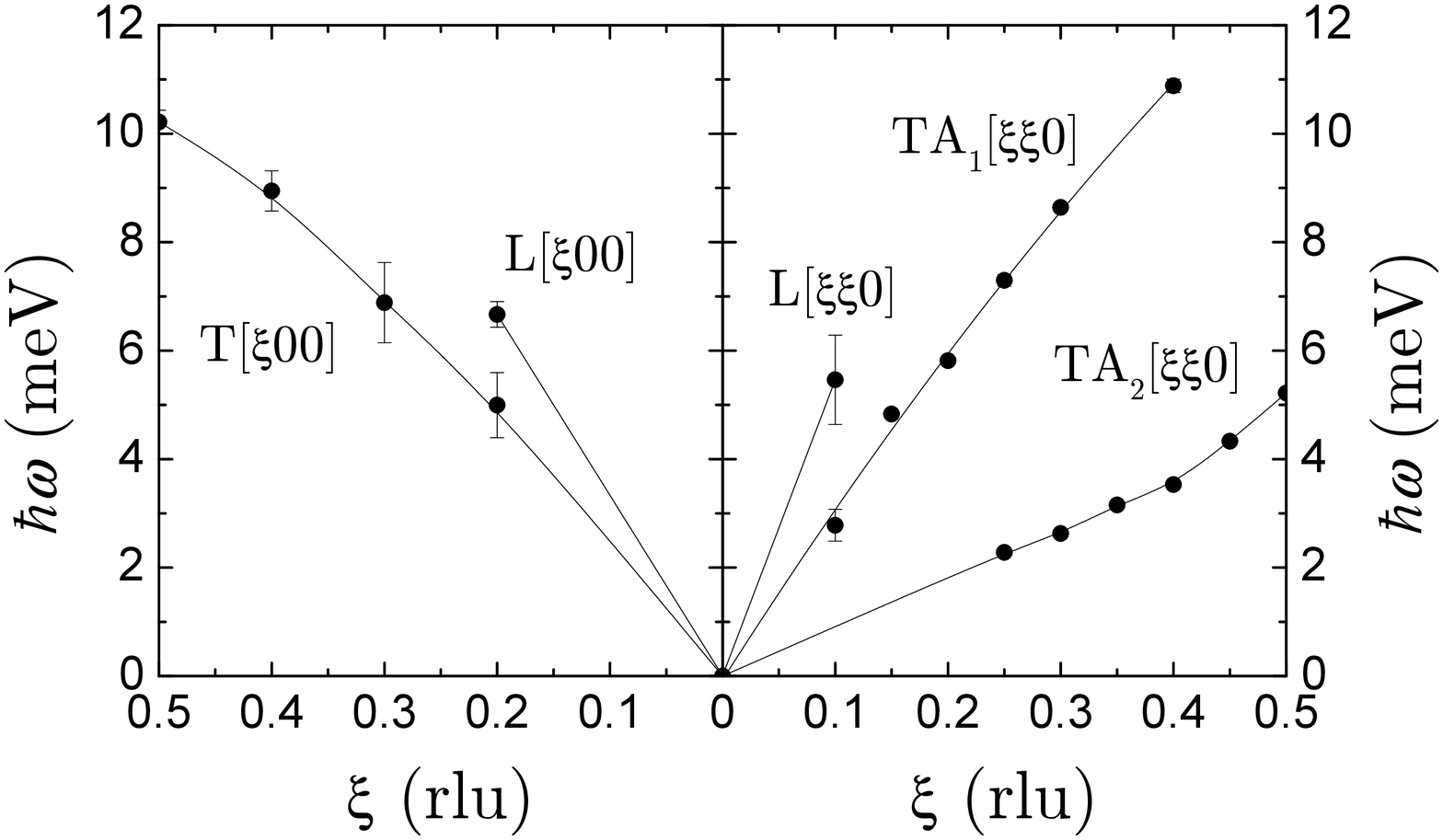}
\caption{Acoustic phonon dispersion curves along the high-symmetry
directions $[\xi 00]$ and $[\xi\xi 0]$ measured at 520 K. Solid
lines are guides to the eye.} \label{fig2}
\end{figure}

We carried out calorimetric measurements in order to characterize
the samples to be studied in the neutron scattering and ultrasonic
experiments.

Calorimetric studies were carried out on small pieces of samples
cut using a low speed diamond saw from the top and bottom of the
large ingot ($\sim 7$ cm long), which was used in the neutron
scattering experiments.  For the sample cut from the top of the
ingot [Fig. \ref{fig1}(a)], calorimetric studies revealed both a
ferromagnetic transition at the Curie temperature $T_{C}=312$ K
and a martensitic transformation at a lower temperature with
characteristic temperature $(M_{s}+A_{f})/2 \approx 297$ K ($M_s$:
martensite-start; $A_f$: austenite-finish). Measurements for the
sample cut from the bottom yielded similar results with a shift of
the transition temperatures of approximately 10 K to higher values
$-$ the shift of the structural transitions being slightly larger
than the shift of the Curie point. These discrepancies are
ascribed to small changes on composition through the length of the
ingot.

Similar measurements carried out in the parallelepiped specimen
used in the ultrasonic experiments revealed a ferromagnetic
transition at $T_{C}=310$ K and a martensitic transformation at
lower temperatures with characteristic temperature
$(M_{s}+A_{f})/2 \approx 161$ K [Fig. \ref{fig1}(b)].

\subsection{\label{subsec:PhononDispersion}Phonon dispersion}

Figure \ref{fig2} shows the phonon dispersion curves determined
from inelastic neutron scattering at 520 K along the high-symmetry
directions $[\xi 0 0]$ and $[\xi \xi 0]$. Note that the plot of
dispersion curves is restricted to half of the Brillouin zone
scheme. The measured phonon spectrum shows the features typical of
bcc materials that undergo martensitic transformations, i.e., low
energies of the phonons of the transverse TA$_{2}[\xi \xi 0]$
branch and a wiggle at $\xi_{0} \approx 0.33$. A similar behavior
has also been reported for the related systems Ni-Mn-Ga and
Ni-Mn-Al with compositions close to stoichiometry
\cite{Zheludev95,Zheludev96,Stuhr97,Manosa01,Moya2006a}. The
existence of such an anomaly in the transverse TA$_{2}$ branch is
ascribed to a strong electron-phonon coupling and the Kohn anomaly
\cite{Lee02}.

\begin{figure}
\includegraphics[width=0.7\linewidth,clip=]{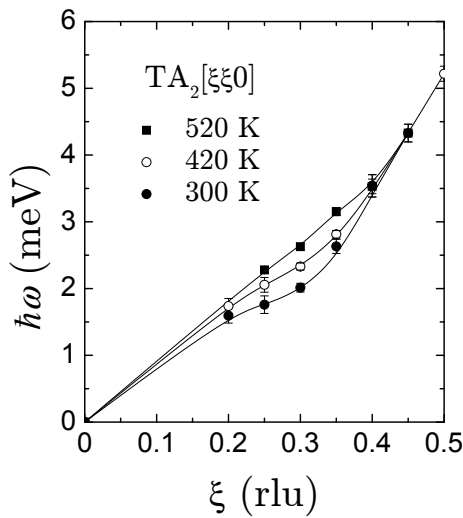}
\caption{Temperature dependence of the TA$_2$[$\xi\xi 0$] branch.
The wiggle at $\xi_{0}\approx 0.33$ deepens with decreasing
temperature thus reflecting the softening of the anomalous phonon.
Solid lines are guides to the eye.} \label{fig3}
\end{figure}

In order to study the temperature dependence of the anomaly
observed in the transverse TA$_{2}$ branch, we measured this
dispersion curve at several temperatures down to room temperature,
i.e., just above the structural transition. The results are shown
in Figure \ref{fig3}. For sake of clarity, measurements only at
three temperatures are presented. As can be seen from the figure,
as the temperature is reduced, the energy of the branch close to
the anomalous phonon decreases, thus reflecting the dynamical
instability of the cubic lattice against distortions of $\{ 110\}$
planes along $\langle 1\bar{1}0 \rangle$ directions. It should be
noted that despite of the fact that the sample transforms
martensitically near room temperature, the wiggle present at high
temperatures does not develop into a marked dip even at 300 K.
This behaviour differs from that observed in Ni-Mn-Ga alloys
\cite{Zheludev95,Zheludev96,Stuhr97,Manosa01} but is similar to
that previously reported in a Ni-Mn-Al alloy close to
stoichiometry \cite{Moya2006a}.

\begin{figure}
\includegraphics[width=0.9\linewidth,clip=]{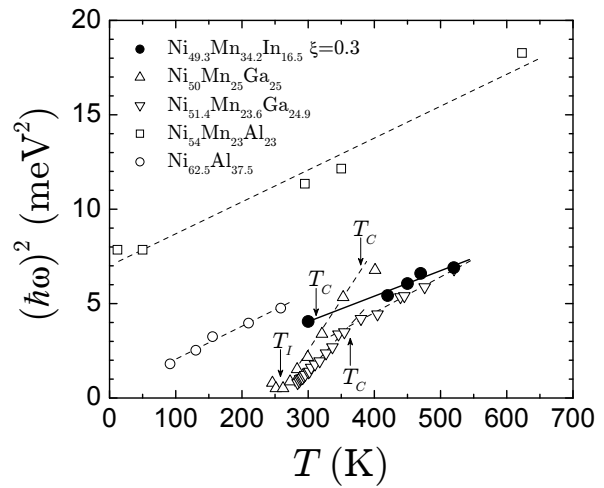}
\caption{Temperature dependence of the energy
squared of the TA$_2$[$\xi\xi 0$] phonon modes close to the
anomalous phonon, $\xi=0.3$ (filled circles). Data from
related systems Ni$_2$MnGa, Ni$_{54}$Mn$_{23}$Al$_{23}$ and
Ni$_{62.5}$Al$_{37.5}$ are also shown for comparison. $T_I$
represents the premartensitic transition temperature. The data
for the latter systems were taken  from references
\cite{Zheludev96}, \cite{Stuhr97}, \cite{Moya2006a} and
\cite{Shapiro91} for Ni$_{50}$Mn$_{25}$Ga$_{25}$,
Ni$_{51.4}$Mn$_{23.6}$Ga$_{24.9}$, Ni$_{54}$Mn$_{23}$Al$_{23}$ and
Ni$_{62.5}$Al$_{37.5}$, respectively. Lines are linear fits to the experimental
data.} \label{fig4}
\end{figure}

In order to compare the behaviour of the different Ni-Mn based
systems, we have plotted in Fig. \ref{fig4} the energy squared of
the anomalous phonons as a function of temperature for different
compounds. Additionally, data for the soft phonon in
Ni$_{62.5}$Al$_{37.5}$ alloy are also plotted. As can be seen from
the figure, the degree of softening in the studied sample is
similar to that of Ni$_{62.5}$Al$_{37.5}$ and
Ni$_{54}$Mn$_{23}$Al$_{23}$, although the energy values of the
latter are higher, which is consistent with the fact that the
Ni-Mn-Al sample does not transform martensitically within the
studied temperature range. By contrast, Ni$_{2}$MnGa alloys show a
more complex behaviour. While the degree of softening in the
paramagnetic state is similar to the studied sample and to the
other related alloys, the softening below the Curie point is much
stronger. This is due to strong spin-phonon coupling exhibited by
these compounds in the ferromagnetic state \cite{Planes1997,Uijttewaal}. It
should be noted
that despite the studied Ni-Mn-In sample is also ferromagnetic,
changes in the degree of softening below $T_{C}$ were not
discernible due to the closeness of the magnetic and the
structural transitions.

The measured dispersion curves are in good agreement with those
obtained from \textit{ab initio} calculations for the
$[110]$ direction in Ni$_{2}$MnIn
\cite{Zayak05}, except for the low energy transverse TA$_{2}$
branch, which exhibits complete softening in the range between
$\xi =0.25$ and $\xi =0.45$. Experimental data are in agreement
with such an instability and show that the minimum is located at
$\xi_{0}\approx 0.33$. However, it can be seen from Fig.
\ref{fig3} that the softening is not complete, i.e., the phonon
frequency remains finite even at the lowest temperatures.

As mentioned above, the large shift of the structural martensitic
transition with the applied magnetic field in these compounds
enables the phase transition to be induced by applying a
magnetic field \cite{Krenke2007}. The microscopic origin of such
strong dependence of the transition temperature on the magnetic
field is ascribed to a strong magnetoelastic coupling, which is responsible
for the change in the relative stability of the martensitic and
cubic phases when the field is applied. Actually, recent \emph{ab
initio} calculations at 0 K for cubic Ni$_{2}$MnIn have shown that
increasing the magnetization due to an external magnetic field
leads to a gradual vanishing of the phonon instability due to the
coupling between vibrational and magnetic degrees of freedom
\cite{Entel2008}. We measured the transverse phonon branch
TA$_{2}$ in the region close to the anomalous phonon at several
applied magnetic fields and at temperatures slightly above the
structural transition. Our results show no significant changes
of the phonon energy with the applied field, so that for these applied fields,
the results cannot confirm
the \emph{ab initio} predictions. However,
it should be noted that the magnetic fields involved in the
calculations are considerably larger than the fields attainable
experimentally in our measurements \cite{MarkusPrivateComm}.
Moreover, it should be also noted that the close proximity of the
magnetic and structural transitions in
Ni$_{49.3}$Mn$_{34.2}$In$_{16.5}$
could inhibit the development of the necessary strong
magnetoelastic coupling before the martensitic transformation
takes place and, hence, prevent the observation of the predicted
magnetic field induced behaviour.

\subsection{\label{subsec:ElasticConstants}Elastic constants}

\begin{figure}
\includegraphics[width=0.7\linewidth,clip=]{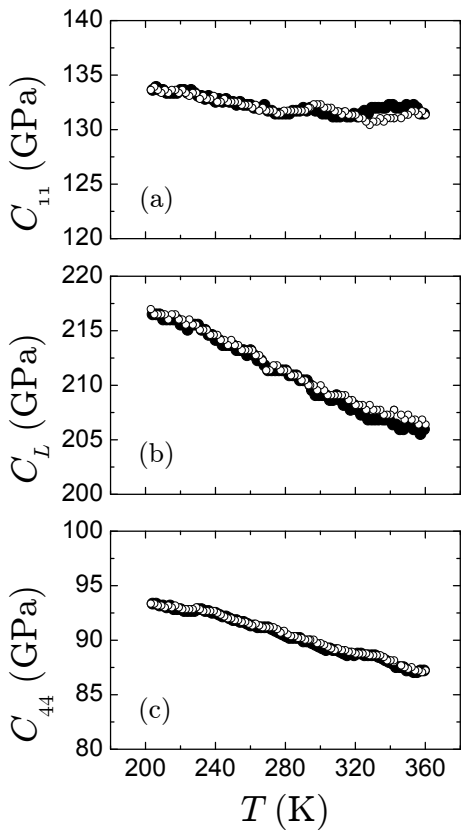}
\caption{Temperature dependence of the elastic moduli (a)
$C_{11}$, (b) $C_{L}=(C_{11}+C_{12}+2C_{44})/2$, and (c) $C_{44}$. Open and
solid symbols
stands for cooling and heating runs, respectively.}
\label{fig5}
\end{figure}

In order to complement the study in the long-wavelength range, we
have measured the elastic constants using the pulse-echo
technique. The results of the temperature-dependent elastic
constants measurements are shown in Fig. \ref{fig5}, where the
thermal behaviour of the three independent elastic constants
$C_{11}$, $C_{L}$, and $C_{44}$ is shown. These are computed from
ultrasonic waves propagating along the $[001]$ and $[110]$
directions down to temperatures close to the martensitic
transformation. The data correspond to cooling (open circles) and
heating (solid circles) runs and are obtained as an average over
several independent runs. As can be seen from the figure, all
three elastic constants increase with decreasing temperature
reflecting the stiffening of the lattice as the temperature is
lowered. $C_{11}$ depends weakly on temperature, similar to the
behaviour reported for the related systems Ni-Mn-Ga
\cite{Stipcich04} and Ni-Mn-Al \cite{Moya2006b}. Additionally, it
should be noted that the elastic moduli show no significant
changes at the Curie point, $T_{C}=310$ K. This behaviour agrees
with that reported in the ferromagnetic Ni-Mn-Ga alloys
\cite{Stipcich04} but differs from the behaviour displayed by the
antiferromagnetic Ni-Mn-Al system. In the latter case,
antiferromagnetic ordering leads to a decrease of the elastic
constants \cite{Moya2006b}. We also note that similar features
have been observed in other antiferromagnetic systems such as
chromium \cite{Muir87}, Fe-Mn and Co-Mn
\cite{Cankurtaran93,Kawald94} below their N\'{e}el temperatures.
Thus, the type of magnetic order developed in the system appears
to influence the elastic properties, although more studies are
required in order to clarify this point.

\begin{figure}
\includegraphics[width=0.7\linewidth,clip=]{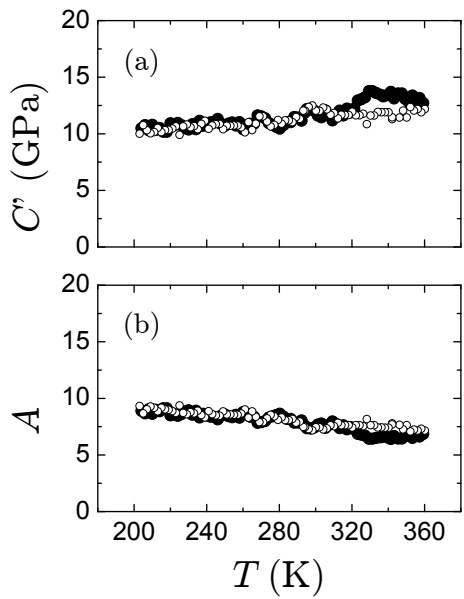}
\caption{(a) Temperature dependence of the shear elastic constant
$C'$ ($=C_{11}+C_{44}-C_{L}$) and (b) the elastic anisotropy $A=C_{44}/C'$
calculated from
the complete set of measured elastic constants (shown in figure
\ref{fig5}). Open and solid symbols
stands for cooling and heating runs, respectively.} \label{fig6}
\end{figure}

From the complete set of measured elastic constants shown in Fig.
\ref{fig5} we can compute the temperature dependence of other
relevant elastic moduli. Fig. \ref{fig6}(a) shows the temperature
dependence of the shear elastic constant $C'$ computed as
$C_{11}+C_{44}-C_{L}$. Owing to the strong attenuation of the
shear waves associated with this mode, it was not possible to
obtain reliable echoes and therefore to measure $C'$ directly. As
can be seen from the figure, $C'$ exhibits a low value and softens
with decreasing temperature. Again, these features reflect the
dynamical instability of the cubic lattice against shearing of $\{
110 \}$ planes along $\langle 1\bar{1} 0 \rangle$ directions.
Additionally, Fig. \ref{fig6}(b) shows the temperature dependence
of the elastic anisotropy calculated as $A=C_{44}/C'$. The elastic
anisotropy at room temperature is similar to those reported for
the related systems Ni-Mn-Ga \cite{Stipcich04} and Ni-Mn-Al
\cite{Moya2006b}, but significantly lower than for Cu-based shape
memory alloys \cite{Planes96}.

\begin{table}\caption{\label{tab:Elastic Constants} Elastic constants obtained
both from ultrasonic methods at 300 K and from the
initial slopes of the acoustic phonon branches $(\xi \rightarrow
0)$ at 520 K. Values determined experimentally show the associated
uncertainty. The calculated values, corresponding to the
stoichiometric Ni$_2$MnIn sample, were estimated from the phonon
branches obtained from \textit{ab-initio} calculations (reference
\cite{Zayak05}).}
\begin{ruledtabular}
\begin{tabular}{ccccc}
& & \text{Ultrasounds} & \text{Neutrons} & \text{\textit{Ab-initio}}\\
& & \text{(GPa)} & \text{(GPa)} & \text{(GPa)}\\
\hline
& $C_{11}$ & $132 \pm 8$ & $190 \pm 15$ & $121$\\
& $C_{L}$ & $210 \pm 10$ & $260 \pm 50$ & $207$\\
& $C_{44}$ & $90 \pm 3$ & $90 \pm 15$ & $101$\\
& $C^{'}$ & $12$ & $12 \pm 4$  & $15$\\
\end{tabular}
\end{ruledtabular}
\end{table}

Up to now, we have discussed the behaviour of the elastic
constants of the Ni-Mn-In alloy obtained from ultrasonic methods.
Elastic constants can be also derived from the initial slope ($\xi
\rightarrow 0$) of the acoustic phonon branches reported in
section \ref{subsec:PhononDispersion}. Table \ref{tab:Elastic
Constants} summarizes the elastic constants obtained from
both experiments. As can be seen, both methods agree well on the values of
$C_{44}$ and $C'$.
There is less agreement on the values for the longitudinal constants $C_{11}$
and
$C_{L}$, which correspond to the initial slope of the $L[\xi 00]$
and $L[\xi \xi 0]$ branches, respectively, for which the values
obtained from the dispersion branches are affected by a large
error due to limited experimental data. Additionally, our
experimental results are in good agreement with elastic constants
obtained from first-principles calculations in the stoichiometric
Ni$_{2}$MnIn \cite{Zayak05}. Again, discrepancies may arise
from the definition of the initial slope, but note that these can also arise
from changes in sample composition and temperature.

\begin{figure}
\includegraphics[width=0.7\linewidth,clip=]{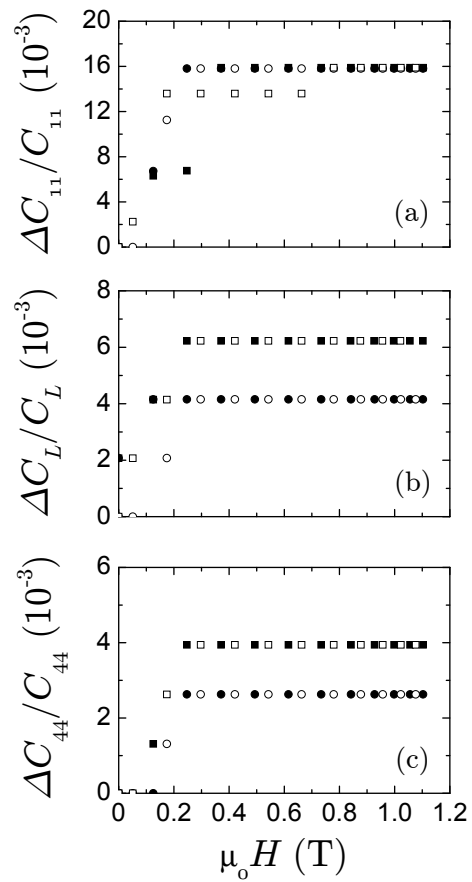}
\caption{Magnetic-field dependence of the relative
change of the elastic constants with respect to the value at zero
field at $T=260$ K. (a) $C_{11}$ versus applied magnetic field
along $[110]$ (circle symbols) and $[1\bar{1}0]$ (square
symbols). (b) $C_{L}$ versus applied magnetic field along $[001]$
(circle symbols) and $[110]$ (square symbols). (c)
$C_{44}$ versus applied magnetic field along $[001]$ (circle
symbols) and $[1\bar{1}0]$ (square symbols). Solid and open
symbols correspond to increasing and decreasing magnetic field
runs, respectively.}
\label{fig7}
\end{figure}

Finally, to investigate further the interplay between elastic and
magnetic properties, we measured the elastic constants as a
function of applied magnetic field at constant temperature.
Results presented in Fig. \ref{fig7} show the
magnetic-field dependence of the relative change of the elastic
constants at $T=260$ K with respect to the zero-field value. The
temperature is located well below the Curie point of the studied sample,
$T_{C}=310$ K. Note that in this case, the ferromagnetic and the
structural transitions are well separated [Fig. \ref{fig1}(b)]
and thus significant magnetoelastic coupling is expected in
contrast to the case for the sample studied by means of neutron scattering
experiments (Sec. \ref{subsec:PhononDispersion}). As can be
seen from the figure, all elastic constants increase up to a
saturation value with increasing magnetic field. This behaviour is
similar to that reported for the prototypical ferromagnetic shape
memory alloy Ni-Mn-Ga \cite{GonzalezComas99} but differs from the
observed behaviour in the antiferromagnetic Ni-Mn-Al system, which
shows an anisotropic magnetoelastic coupling \cite{Moya2006b}. As
discussed before, the type of magnetic order present in the
different compounds of the Ni-Mn based family
influences their elastic properties, as reflected by the different
temperature dependence of the elastic constants across the
magnetic transition \cite{Stipcich04,Moya2006b}. Additionally, the magnetic
field dependence of the elastic moduli further evidences the
dependence of the elastic properties on the type of magnetic
ordering indicating that the magnetic order also determines the
magnetoelastic response of these alloy systems
\cite{GonzalezComas99,Moya2006b}.

\section{Summary and conclusions}
\label{sec:Summary}

We have carried out neutron scattering and ultrasonic experiments
in order to study in detail the lattice dynamics of Ni-Mn-In
Heusler alloys with composition close to the
Ni$_{50}$Mn$_{34}$In$_{16}$ magnetic superelastic alloy. In order
to study the interplay between dynamical and magnetic degrees of
freedom, we performed both kind of experiments in applied magnetic
fields. The most relevant outcomes from this study are:

- The values obtained for the elastic constants from both neutron
scattering and ultrasonic experiments are in good agreement with
each other. In addition, data reported for the stoichiometric
Ni$_{2}$MnIn from \emph{ab initio} calculations are consistent
with the measured values.

- The TA$_2$ phonon branch exhibits a wiggle in the vicinity of
$\xi_{0}\approx 1/3$ at temperatures well above the Curie point.
Upon cooling this anomaly is enhanced, but even at temperatures
close to the martensitic transition the dip is less pronounced
than in Ni-Mn-Ga.

- The elastic constant $C'$ also softens with decreasing
temperature down to the martensitic transition. Both the anomalous
phonon energy and $C'$ have finite values at the transition
temperature. The monotonic decrease in the anomalous phonon energy
and in the elastic constant indicate the absence of a
premartensitic phase (associated with the condensation of the
anomalous phonon) in this alloy system.

- Within experimental errors, the development of ferromagnetic
order at the Curie point does not modify the rate of softening.

- While no magnetic field dependence of phonon energies has been
measured for fields up to 4 T, all elastic constants increase with
increasing magnetic field thus evidence the existence of
magnetoelastic coupling at the long-wavelength limit.

\begin{acknowledgments}

This work received financial support from the CICyT (Spain),
Project No. MAT2007-62100 and from the Deutsche
Forschungsgemeinschaft (SPP 1239). XM
acknowledges support from Comissionat per a Universitats i Recerca
(CUR) del Departament d'Innovaci\'{o}, Universitats i Empresa de
la Generalitat de Catalunya. Experiments at Oak Ridge National
Laboratory's High Flux Isotope Reactor was sponsored by the
Scientific User Facilities Division, Office of Basic Energy
Sciences, U. S. Department of Energy. Work at the Ames Laboratory was supported
by the Office of Basic Energy Sciences USDOE, under Contract No.
DE-AC02-07CH11358.

\end{acknowledgments}


\bibliography{apssamp}

\end{document}